\documentclass[conference]{IEEEtran}
\IEEEoverridecommandlockouts
\usepackage{cite}
\usepackage{subfigure}
\usepackage{amsmath,amssymb,amsfonts}
\usepackage{algorithmic}
\usepackage{graphicx}
\usepackage{textcomp}
\usepackage{xcolor}
\usepackage{multirow}

\def\BibTeX{{\rm B\kern-.05em{\sc i\kern-.025em b}\kern-.08em
    T\kern-.1667em\lower.7ex\hbox{E}\kern-.125emX}}
\begin{document}

\title{ColonNet: A hybrid of DenseNET121 \& U-NET model for detection and Segmentation of GI Bleeding}

\author{
\IEEEauthorblockN{Ayushman Singh}
\IEEEauthorblockA{\textit{CSE} \\
\textit{IIIT Ranchi, India}\\
ayushmaansingh72@gmail.com}
\and
\IEEEauthorblockN{Sharad Prakash}
\IEEEauthorblockA{\textit{CSE} \\
\textit{IIIT Ranchi,India}\\
sharadprakash117@gmail.com}
\and
\IEEEauthorblockN{Aniket Das}
\IEEEauthorblockA{\textit{CSE} \\
\textit{IIIT Ranchi,India}\\
aniket.das1203@gmail.com}
\and
\IEEEauthorblockN{Nidhi Kushwaha}
\IEEEauthorblockA{\textit{Faculty} \\
\textit{IIIT Ranchi,India}\\
nidhi@iiitranchi.ac.in}
}
\maketitle

\begin{abstract}
This study presents an integrated deep learning model for automatic detection and classification of Gastrointestinal bleeding in the frames extracted from Wireless Capsule Endoscopy (WCE) videos. The dataset has been released as part of Auto-WCBleedGen Challenge Version V2 hosted by the MISAHUB team. Our model attained the highest performance among 75 teams that took part in this competition. It aims to efficiently utilizes CNN based model i.e. DenseNet and UNet to detect and segment bleeding and non-bleeding areas in the real-world complex dataset. The model achieves an impressive overall accuracy of 80\% which would surely help a skilled doctor to carry out further diagnostics.

\end{abstract}


\section{Introduction}

Gastrointestinal (GI) bleeding is a pathological condition characterized by hemorrhaging within the digestive tract. The influx of blood into the GI tract poses a spectrum of complications, ranging from acute risks to chronic ramifications\cite{Nakashima2018},\cite{Gralnek2015}. According to estimates by the World Health Organization (WHO), GI bleeding contributes to approximately 300,000 deaths globally each year. Over the past decade, advancements in diagnostic technologies, such as Wireless Capsule Endoscopy (WCE)  \cite{Bordbar2023},\cite{B2022},\cite{Jensen2024} have significantly enhanced our understanding of GI bleeding within the gastrointestinal (GI) tract. WCE involves the recording of a video depicting the trajectory of the GI tract using a wearable device over a duration of 8–12 hours, generating between 57,000–100,000 frames of footage. Presently, the process of reviewing a single patient's recorded WCE video via frame-by-frame analysis requires approximately two to three hours of meticulous examination by a skilled gastroenterologist. However, this method is labor-intensive and susceptible to human error. Given the global shortage of gastroenterologists, there is a pressing need for research aimed at developing state-of-the-art Artificial Intelligence (AI) models that are reliable, interpretable, and widely applicable \cite{Saito2020} \cite{MarinSantos2022}.

\section{Dataset Description}
The MISAHUB team provided both training and testing datasets for the version 2 challenge \cite{palakbleedingtrain} in February 2024. The training dataset comprises 2,618 WCE frames, encompassing instances of bleeding and non - bleeding, sourced from various internet repositories. The test set provided later on was divided into two groups: Test set 1 had 49 images with very subtle and in certain cases almost imperceptible instances of bleeding. Test set 2 had 515 images with varying sizes of bleeding segments.

\section{Preprocessing of Dataset}
In medical image analysis, achieving deformation and rotation invariance is critical for reliable outputs. To enhance our model's robustness, we used data augmentation techniques on the train dataset from MISAHUB, splitting it 80:20 for training and validation. Leveraging the U-Net architecture's adaptability, we employed horizontal and vertical flips, rotations, and resizing. Bounding boxes were included for the Detection path. Horizontal and vertical flips were applied with a 0.5 probability to maintain rotation neutrality.
A few other techniques including CLAHE, Erosion, and Dilation were also used, but only contributed to worsened trained results and were dropped.

\section{Proposed Model}
\begin{figure}[!h]
    \centering
    \includegraphics[width=1\linewidth]{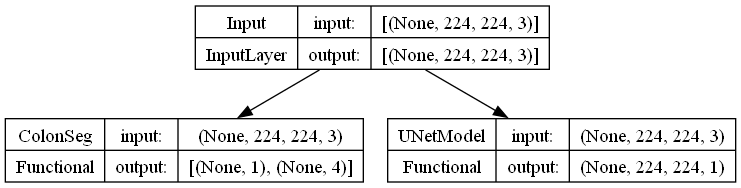}
    \caption{Architecture of Proposed model}
    \label{ColonNet}
\end{figure}

The ColonNet model (as depicted in Fig. \ref{ColonNet}) consists of two branches: ColonSeg and UNetModel, which are detailed as follows.
\subsection{ColonSeg}\label{AA}
\begin{figure}
    \centering
    \includegraphics[width=1\linewidth]{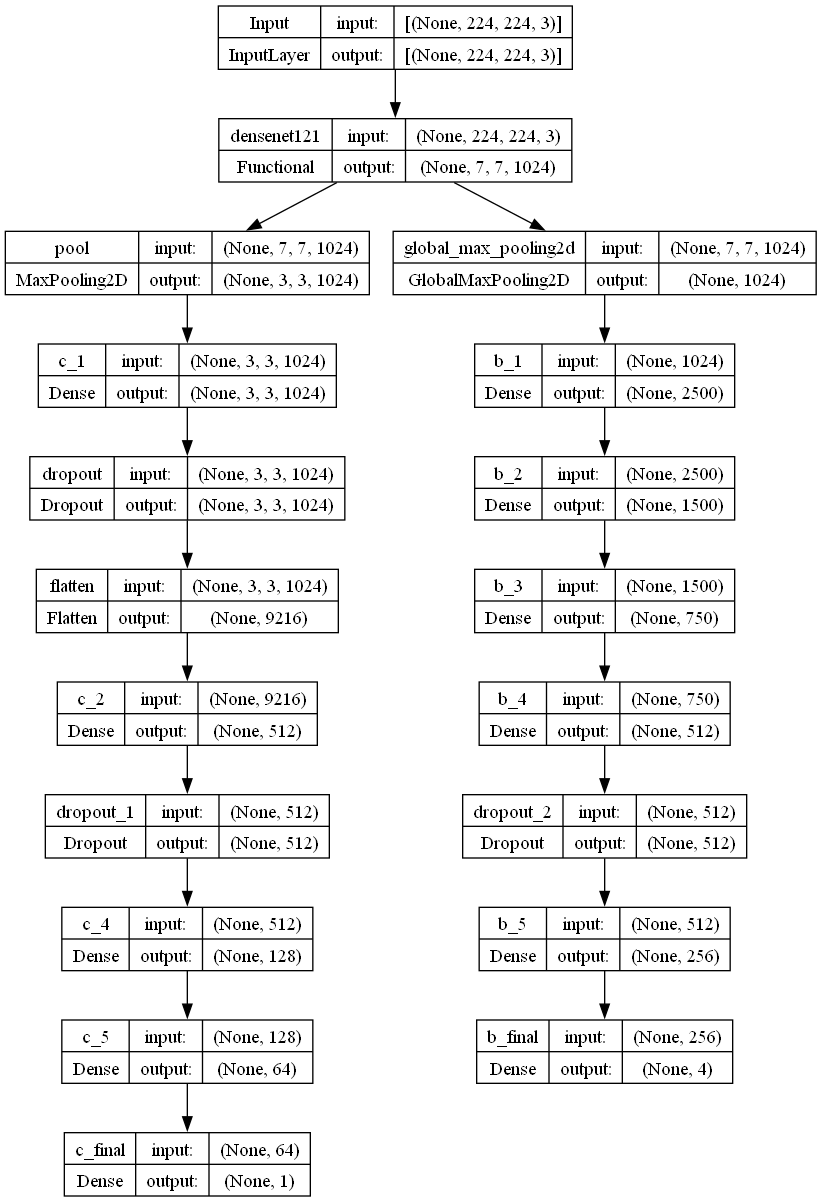}
    \caption{Architecture of ColonSeg}
    \label{ColonSeg}
\end{figure}

The architecture of ColonSeg has been shown in Fig. \ref{ColonSeg}. It is responsible for the classification and detection of the bleeding in the GI tract. An input image of size 224x224x3 is passed through a DenseNet121 block \cite{Cinar2022}, which gives an output of size 7x7x1024. This output is then further passed through two different branches, responsible for classification and detection each. In the classification branch the DenseNet output is passed through multiple Dense layers with RELU activation, and a final sigmoid activated layer for output. The Detection block passes the same DenseNet output through multiple fully connected layers with RELU and ELU activations, and a final sigmoid layer for output.

\subsection{UNetModel}
The Segmentation branch is implemented completely based on the U-Net architecture \cite{U-net} with two different paths, one for downsampling and another for upsampling. The input image is downsampled while increasing the no. of feature channels; this downsampled image is then upsampled along with concatenation with its corresponding downsampled output. The final segmentation mask is received through a sigmoid activated layer.The Detection branch was trained first using the Mean Squared Error loss using only bleeding images. Classification branch was trained on all the images, and the trainable parameters of the DenseNet121 and the Detection branch were frozen. Binary Cross Entropy function was used as the loss function for training.The U-Net branch was trained separately on all the images, and the Focal Teversky loss function \cite{focal-tversky} was used for it.

\section{Experimental Settings}
Training and testing occurred on a Kaggle platform using a GPU P100. UNet trained for 40 epochs, detection for 10 epochs, and classification for 20 epochs, consuming 15, 9, and 2 minutes, respectively, per epoch. Prediction and segmentation on the test dataset took roughly 28 seconds.

\begin{table}[ht]\small
\resizebox{\columnwidth}{!}{%
\begin{tabular}{|c|l|c|c|}
\hline
\textbf{Metric} & \textbf{} & \multicolumn{1}{l|}{\textbf{Test Dataset 1}} & \multicolumn{1}{l|}{\textbf{Test Dataset 2}} \\ \hline
\multirow{3}{*}{\textbf{Classification}} & \textbf{Accuracy}         & 0.4898 & 0.8175 \\ \cline{2-4}
                                         & \textbf{Recall}           & 0.4898 & 0.7448 \\ \cline{2-4}
                                         & \textbf{F1-score}         & 0.8056 & 0.8213 \\ \hline
\multirow{2}{*}{\textbf{Detection}}      & \textbf{Avg. Precision}   & 0.5000 & 0.8250 \\ \cline{2-4}
                                         & \textbf{IoU Score}        & 0.1254 & 0.6315 \\ \hline
\multirow{2}{*}{\textbf{Segmentation}}   & \textbf{Dice-Coefficient} & 0.1767 & 0.6833 \\ \cline{2-4}
                                         & \textbf{IoU Score}        & 0.2029 & 0.6128 \\ \hline
\end{tabular}%
}
\caption{Performance Statistics}
\label{Performance Statistics}
\end{table}

\section{Discussion}
\begin{figure}[!h]
    \centering
    \subfigure[Bounding Box]{\includegraphics[width=0.25\linewidth]{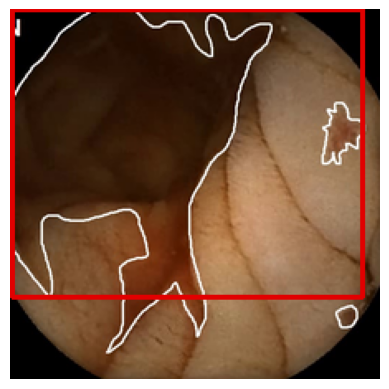}}
    \centering
    \subfigure[Image mask]{\includegraphics[width=0.25\linewidth]{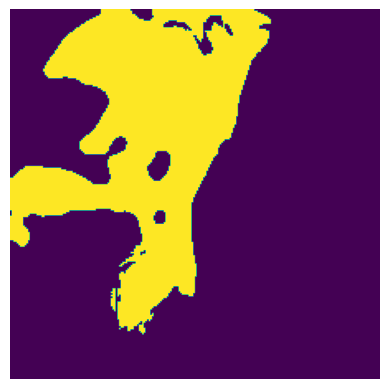}}
    \centering
    \subfigure[CAM Plot]{\includegraphics[width=0.25\linewidth]{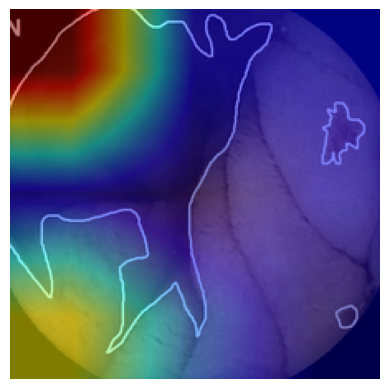}}
    \caption{Result on test dataset 1}
    \label{Test Image 1}
\end{figure}

The final results are summarized in Table \ref{Performance Statistics}. Test dataset 1 mainly contained images with small bleeding patches, yielding a classification accuracy of around 50\%. Fig. \ref{Test Image 1} illustrates one exemplary prediction from the test dataset.

\begin{figure}[!h]
    \centering
    \subfigure[Bounding Box]{\includegraphics[width=0.25\linewidth]{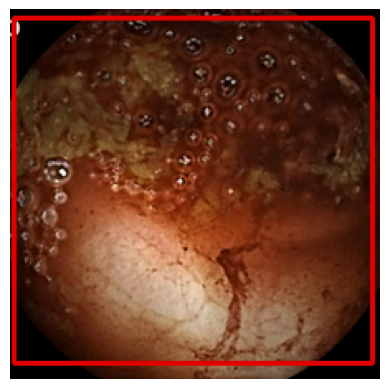}}
    \centering
    \subfigure[Image mask]{\includegraphics[width=0.25\linewidth]{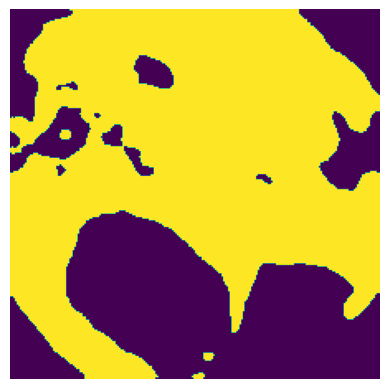}}
    \centering
    \subfigure[CAM Plot]{\includegraphics[width=0.25\linewidth]{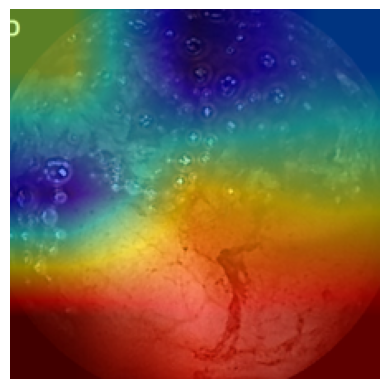}}
    \caption{Result on test dataset 2}
    \label{Test Image 2}
\end{figure}

In contrast, test data set 2 showed a higher variance in image content, including more easily perceptible bleeding instances, resulting in a markedly improved accuracy of 80\%. Moreover, the segmentation's Dice coefficient and Intersection over Union (IoU) score were notably superior in test dataset 2 compared to test dataset 1. Fig. \ref{Test Image 2} shows one of the best results obtained from test dataset 2.\\
We tried to handle the presence of bubbles and convoluted surfaces withing the GI tract using Erosion, Dilation, CLAHE, and other techniques, but this resulted in loss of vital information where subtle bleeding patches were present, resulting in misclassification.\\
We trained the model with 3 separate backbones: VGG19, ResNet, and DenseNet. Although the VGG19 and ResNet models were marginally better at classification task compared to DenseNet, the latter outperformed them in the detection task, with very little deprecation in the other task. This led us to the conclusion that the large feature extraction of the DenseNet model helped in detection task, much better compared to the other two models.

\section{Conclusion and Future Scope}.
The model exhibited commendable performance, achieving classification accuracy 80\% without any pre-processing specific to the data set. Challenges arose from image with noise and bubbles, aggravated by pre-processing. The absence of contextual cues led to misinterpretations, difficulty in detecting smaller bleeding patches, and overlooking multiple bleeding sites, often results in the bounding box of the largest area. 

\section{Acknowledgments}
As participants in the Auto-WCEBleedGen Version V2 Challenge, we fully comply with the competition rules as outlined in \cite{hub2024auto} and the challenge website. Our methods have used the training and test data sets provided in the official release in \cite{palakbleedingtrain} and \cite{palakbleedingtest} to report the results of the challenge.

\bibliographystyle{unsrt}
\bibliography{references}

\end{document}